# A SQUID based read-out of sub-attoNewton force sensor operating at millikelvin temperatures

O. Usenko, A. Vinante, G. H. C. J. Wijts, T.H. Oosterkamp
Leiden Institute of Physics, Leiden University, the Netherlands

An increasing number of experiments require the use of ultrasensitive nanomechanical resonators. Relevant examples are the investigation of quantum effects in mechanical systems [1] or the detection of exceedingly small forces as in Magnetic Resonance Force Microscopy (MRFM) [2]. The force sensitivity of a mechanical resonator is typically limited by thermal fluctuations, which calls for detection methods capable of operating at ultralow temperature. Commonly used interferometric techniques, despite their excellent sensitivity, may not be an optimal choice at millikelvin temperatures, because of unwanted resonator heating caused by photon absorption. Although alternative detection techniques based on microwave cavities [3] [4] [5] have shown to perform better at ultralow temperature, these techniques still suffer from the fact that the detection sensitivity decreases as the power input is decreased.
Here, we present a measurement approach based on the detection, through a Superconducting Quantum Interference Device (SQUID), of the change of magnetic flux induced in a coil by the motion of a magnetic particle attached to a resonator. This detection scheme avoids direct heating of the resonator, as it does not involve reflecting optical or microwave photons to the resonator. By cooling an ultrasoft silicon resonator to 25 mK, we achieve a force noise of 0.5 aN in a 1 Hz bandwidth. We believe this detection technique can in principle be used even at sub-millikelvin temperatures. Furthermore, it could be used to improve the sensitivity of MRFM experiments, which aim at three dimensional imaging at atomic resolution.

Our method involves attaching a magnetic particle to the end of the resonator (Figure 1a), which causes a change of magnetic flux in a coil, positioned close to the resonator, whenever the resonator moves (figure 1b). The silicon resonator consists of a 100 nm thick single crystal beam which is 5 μm wide and 100 μm long, fabricated as reported in [6] with an estimated spring constant of $1.44 \cdot 10^{-4}$ N/m for the lowest flexural mode. We have attached a 4.5 μm micrometer diameter magnetic sphere of a Neodymium based alloy to the end of the resonator using a nano-manipulator inside an electron microscope [7], and magnetized it in a 3 T field at room temperature.
The resonator is then placed at about 10 μm above the edge of a detection coil consisting of a square thin film Nb coil with 22 windings, 670 μm wide and a total inductance of 0.6 μH. The change in flux due to the motion of the sphere is measured using a circuit (Figure 1b) which includes the detection coil, a calibration transformer and the input coil which couples the signal to the SQUID, which measures the change in flux. The whole assembly is cooled in vacuum and thermally anchored to the mixing chamber of a dilution refrigerator with a base temperature of 8 mK. At cryogenic temperature, the resonant frequency is 3084 Hz and the quality factor is weakly dependent on temperature, reaching about 38000 at 11 mK.

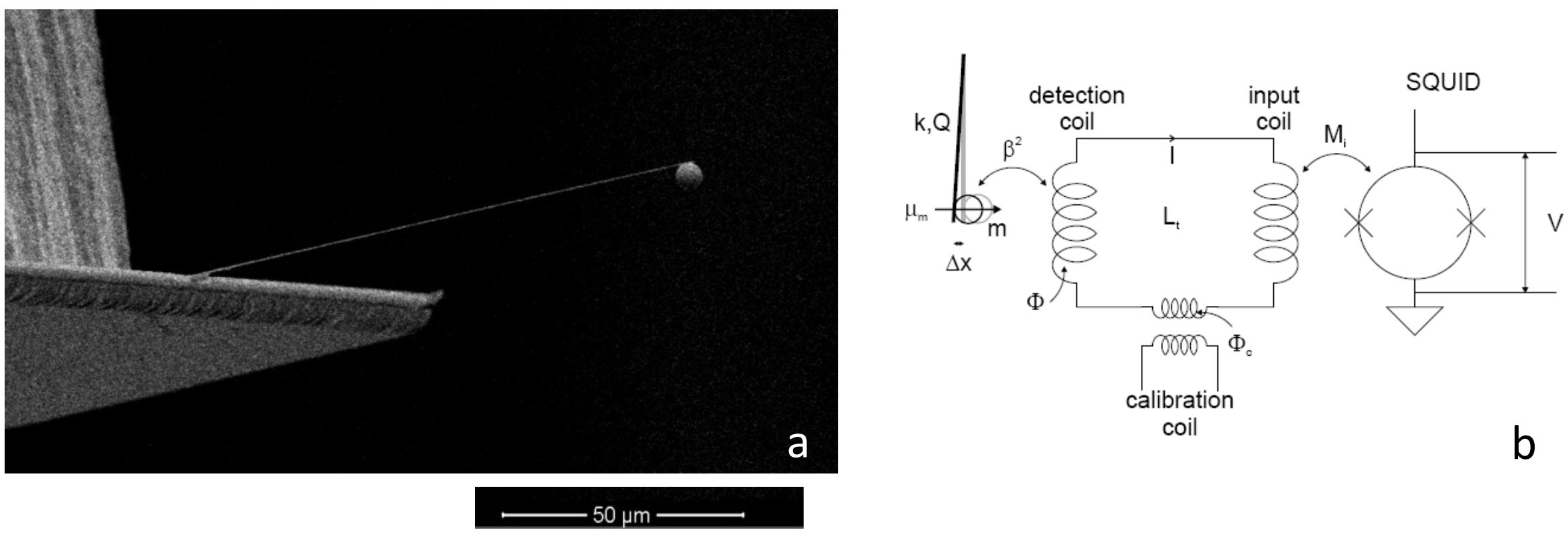

*Figure 1 a) An electron microscopy image of the silicon resonator with a magnetic sphere attached to its end. The single crystal beam is 100 nm thick, 5 micrometer wide and 100 micrometer long. b) circuit diagram illustrating the detection scheme. The motion of the resonator induces a current in the loop which is detected with the SQUID. The coupling between the motion of the resonator and the detection circuit can be calibrated with the calibration transformer.*

To calibrate the detected signal in terms of the resonator energy, we apply a flux to the detection circuit through a calibration transformer. The measured current in the detection circuit as a function of frequency shows a resonant response due to the driving of the mechanical resonator. Accurate measurement of this response allows us to measure the effective coupling between resonator and detection circuit, and to calibrate the energy of the resonator (see supplementary information).

In the inset to figure 2, the spectral density of the SQUID signal due to the thermal motion of the resonator is shown at two separate bath temperatures of 1014 mK and 11 mK.

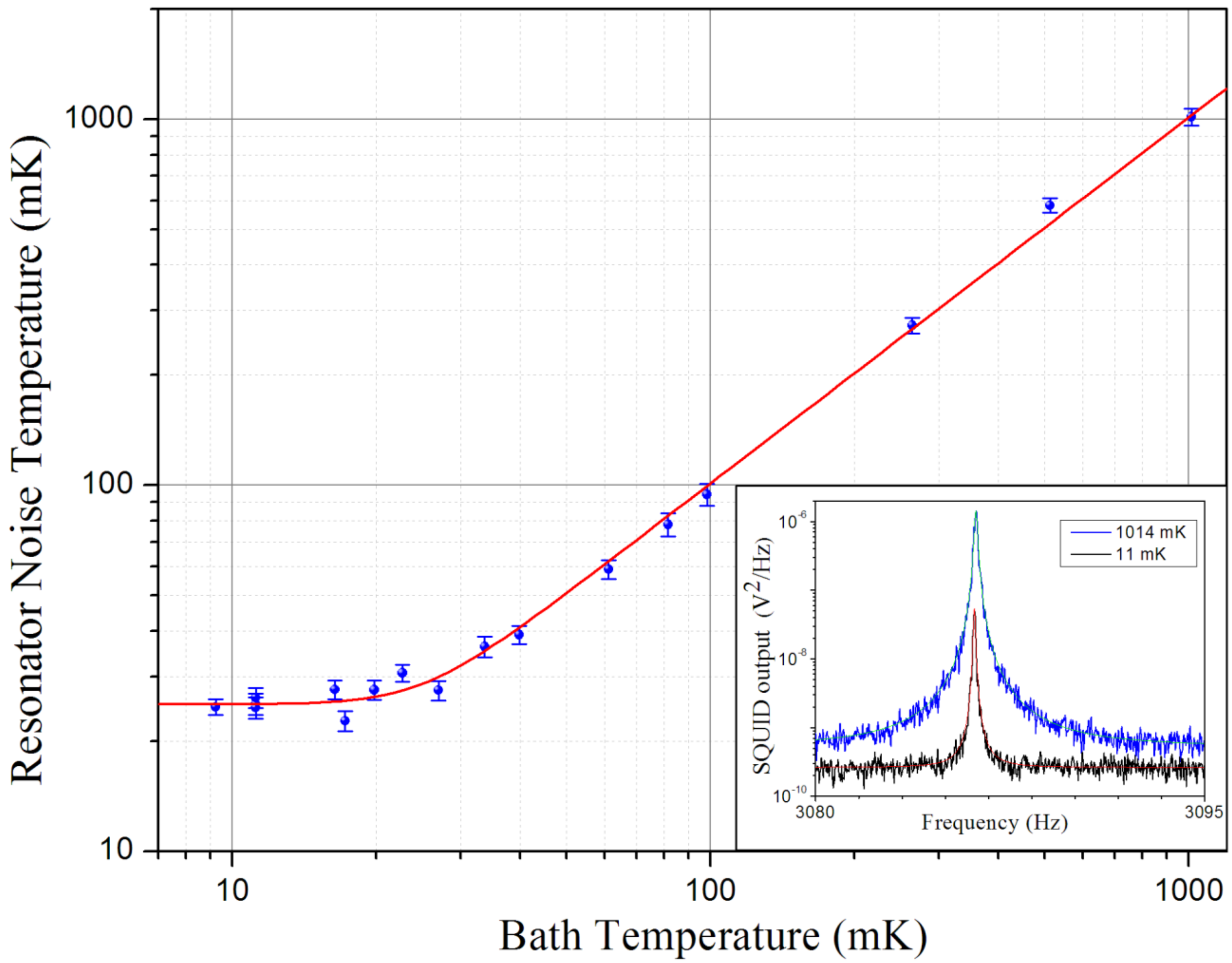

*Figture 2. The temperature of the 1st mode of the resonator (obtained by integrating the thermal noise of the resonator) is plotted as a function of the temperature of the dilution refrigerator. Inset: Frequency dependence of the resonator motion at two different temperatures.*

In the main panel of figure 2 the calibrated thermal energy of the resonator, expressed as effective temperature, is shown as a function of the mixing chamber temperature.

The red line presents a fit to the data of the form $(T^n+T_0^n)^{1/n}$, where $T_0$ is the saturation temperature and the exponent $n$ is determined by the temperature dependence of the limiting thermal resistance which scales as $T^{(n-1)}$ [8]. The fit yields a saturation temperature $T_0=25\pm1$ mK , while the exponent $n=5\pm2$ is consistent with a limiting thermal resistance due to 3D phonons in the chip supporting the resonator [8] or to a thermal boundary resistance.

The force noise power spectral density is given by $S_F = 4k_B T\gamma$, where $\gamma$ is the damping of the resonator, defined as $\gamma = k/(2\pi f_0 Q)$. For our saturation temperature and the measured Q we obtain a noise level of 0.51±0.03 aN/√Hz.

**Materials and methods**

We used a single sphere from a powder of Nd-Pr-Fe-Co-Ti-Zr-B alloy (MQP-S-11-9-20001-070) from Magnequench. We used a two stage SQUID amplifier, made of a commercial Quantum Design sensor SQUID and a custom designed amplifier SQUID [9]. The SQUIDs were operated with commercial direct readout electronics from Star Cryoelectronics.

**Conclusion and discussion**

By assembling a relatively simple experimental setup and cooling our resonator to an effective temperature of 25 mK we achieved a force noise of 0.51±0.03 aN/√Hz. We believe that by further improving the experimental setup, the effective temperature and thus the force noise can be reduced further, possibly to sub-millikelvin temperatures. A force sensor such as the one presented here, operating at an effective temperature of 10 mK would have a force sensitivity of 0.3 aN/sqrt(Hz).

Furthermore, the described technique is also suitable for detecting the motion of resonators based on nanowires which pose problems for use in interferometric detection due to very low reflectivity [10]. For example, single crystal SiC nanowires have demonstrated a damping factor as low as 4 fNm/s [11], which would yield a force noise below 0.1 aN/√Hz at 10 mK.

In MRFM experiment containing a single proton spin flipping in a field gradient of $2\times10^6$ T/m (e.g. at 50 nm from a 1 μm diameter magnet with a magnetization of 0.75 T), a force of 0.03 aN is generated, which would be detectable in an averaging time of less than 500 seconds.